\documentclass[twocolumn]{aastex63}

\received{October 9, 2020}
\accepted{\today}
\shorttitle{Sample article}
\shortauthors{Horiuchi et al.}
\begin{document}

\title{Simultaneous Multicolor Observations of Starlink's Darksat by The Murikabushi Telescope with 
$\it{MITSuME}$}

\author[0000-0001-5925-3350]{Takashi Horiuchi}
\affil{Ishigakijima Astronomical Observatory, \\
Public Relations Center, \\
National Astronomical Observatory of Japan, \\
1024-1 Arakawa, Ishigaki, Okinawa, 907-0024, Japan}

\author[0000-0001-8221-6048]{Hidekazu Hanayama}
\affiliation{Ishigakijima Astronomical Observatory, \\
Public Relations Center, \\
National Astronomical Observatory of Japan, \\
1024-1 Arakawa, Ishigaki, Okinawa, 907-0024, Japan}

\author[0000-0003-2775-7487]{Masatoshi Ohishi}
\affiliation{Spectrum Management Office, Public Relations Center, \\
National Astronomical Observatory of Japan, \\ 
2-21-1 Osawa Mitaka Tokyo 181-8588, Japan}
\affiliation{Department of Astronomical Science, \\
SOKENDAI (The Graduate University for Advanced Studies), \\
2-21-1 Osawa, Mitaka, Tokyo 181-8588, Japan}



\begin{abstract}
We present the SDSS $g'$-, the Cousins $R_{\rm c}$-, and $I_{\rm c}$-band 
magnitudes and associated colors of the Starlink’s STARLINK-1113 (one of the standard 
Starlink satellites) and 1130 (Darksat) with a darkening treatment to its surface. By the 
105 cm Murikabushi telescope/$\it{MITSuME}$, simultaneous multicolor observations 
for the above satellites were conducted four times: on April 10, 2020, May 18, 2020 (for 
Darksat), and June 11, 2020 (for Darksat and STARLINK-1113). We found that (1) the 
SDSS $g'$-band apparent magnitudes of Darksat ($6.95~\pm~0.11-7.65~\pm~0.11$ mag) 
are comparable to or brighter than that of STARLINK-1113 ($7.69~\pm~0.16$ mag), (2) the 
shorter the observed wavelength is, the fainter the satellite magnitudes tend to become, (3) the 
reflected flux by STARLINK-1113 is extremely ($> 1.0$ mag) redder than that of Darksat, (4) 
there is no clear correlation between the solar phase angle and orbital-altitude-scaled magnitude, 
(5) by flux-model fitting of the satellite trails with the blackbody radiation, it is found that the 
albedo of Darksat is about a half of that of STARLINK-1113. 
Especially, the result (1) is inconsistent with the previous studies. However, considering 
both solar and observer phase angles and atmospheric extinction, the brightness of  
STARLINK-1113 can be drastically reduced in the SDSS $g'$ and the Cousins 
$R_{\rm c}$ band. Simultaneous multicolor-multispot observations more than three colors 
would give us more detailed information regarding the impact by the low Earth orbit satellite 
constellations.  
\end{abstract}


\section{Introduction} \label{sec:intro}

SpaceX launched 60 Starlink satellites to low Earth orbit (LEO) on May 24, 2019 as the 
first batch of a large constellation. Furthermore, SpaceX plans to launch 42,000 Starlink LEO 
communication satellites in total until Mid 2020s. On June 3, 2019, the International 
Astronomical Union (IAU) expressed its concern on the fact that the extremely bright 
magnitude of these communication satellites affect astronomical observations and the 
pristine appearance of the night sky\footnote{https://www.iau.org/news/pressreleases/detail/iau2001/}. 
In response to the concern by the IAU, SpaceX has tried to reduce satellite brightness, 
and has also asked astronomers to measure the satellite brightness. After the launch of the 
Starlink satellites, their apparent magnitude and impact on astronomical 
observations were reported by previous studies using ground-based telescopes. 
\citet{2020A&A...636A.121H} investigated the impact by mega-constellations of the LEO satellites 
in the optical and IR wavelength regions on the ESO telescopes. They concluded that 
very wide-field imaging surveys will be terribly ruined due to saturation and/or ghosting 
by a satellite. \citet{2020ApJ...892L..36M} also concluded that certain types of observation 
such as long-exposure and twilight observations with wide fields of view (FoV) will be 
significantly affected by the Starlink satellites. 
 
SpaceX launched the third batch of the 60 LEO satellites on January 7, 2020. One of the 60 
satellites is the prototype satellite, STARLINK-1130 (nicknamed Darksat) where the 
communication antenna is coated with painting to reduce reflected sunlight to the 
Earth \citep[see Figure 7 of ][]{2020arXiv200612417T}. As the first observational 
attempt, \citet{2020A&A...637L...1T} estimated the Sloan $g'$ magnitude of 
Darksat and STARLINK-1113 and suggested that Darksat is 
0.77 $\pm$ 0.05 magnitude fainter than STARLINK-1113. However, simultaneous multicolor 
observations for these satellites have not been reported yet despite the importance of this 
kind of observation. The simultaneity of the observations ensures the same conditions  
in multiple bands: same coordinate, airmass, and exposure time, and so on. 
Not only that, it is able to examine the darkening effects of satellite-surface 
by the color estimation and/or radiation model fitting for these LEO communication 
satellites under the simultaneous multicolor observations.

In this paper, we report the multi-band (the SDSS $g'$, the Cousins $R_{\rm c}$, 
and $I_{\rm c}$; hereafter $g', R_{\rm c}$, and $I_{\rm c}$ for simplicity) magnitudes 
and colors of Darksat and STARLINK-1113 with simultaneous multicolor observations 
using the 1.05m Murikabushi telescope/$\it{MITSuME}$ system. In Section 2, the 
observations and the data analysis for Darksat and STARLINK-1113 are presented. 
We show the apparent and orbital-altitude-scaled magnitudes of the satellites in Section 3. 
In Section 4, we discuss (1) the effects of atmospheric extinction to the satellites, 
and (2) modeling the AB flux of the satellites. 

\section{Observations and Data Analysis} \label{sec:floats}
\subsection{Observations}
Ishigakijima Astronomical Observatory (IAO) of the National Astronomical Observatory of 
Japan operates the 105 cm Murikabushi telescope (F12) 
with $\it{MITSuME}$\footnote{This is an abbreviation of Multi-color Imaging Telescopes for 
Surveys and Monstrous Explosions. It means “three eyes" in Japanese.}. 
The $\it{MITSuME}$ system equipped with the Murikabushi telescope has three CCD 
cameras (Apogee Alta U-6) with 1024 $\times$ 1024 pixels, and enables us 
simultaneous $g'~$(477~nm)-, $R_{\rm c}~$(649.2~nm)- and $I_{\rm c}~$(802~nm)-band 
observations, where the values in the parenthesis indicate the effective wavelengths in 
each band. Among the three bands $R_{\rm c}$ band has the most highest sensitivity. 
The F-value of the Murikabushi telescope is adjusted from F12 to F6.5 via the conversion 
lenses in the $\it{MITSuME}$ system; it yields a FoV of 12.3 $\times$ 12.3 arcmin$^2$ 
and a pixel scale of 0.72 arcsec pixel$^{-1}$. 

Based on the two-line element (TLE) data from the satellite catalog of the Celestrak 
website\footnote{https://celestrak.com/satcat/search.php} and the forecast time and equatorial 
coordinates extracted from 
$\tt{Heavensat}$\footnote{http://www.sat.belastro.net/heavensat.ru/english/index.html}, 
we carried out simultaneous multicolor observations for the light trail of Darksat and 
STARLINK-1113 using the Murikabushi telescope/$\it{MITSuME}$ between April and 
June, 2020. Table 1 lists the observation log in this study. Since the satellites move so fast, 
we were not able to track them. Instead we pointed the telescope to the calculated position 
of a satellite, and waited for the satellite passing by the field of view.

The observed flux of satellites, $f_{\rm sat}$, is inversely proportional to angular  
velocity of the satellite along the celestial sphere, $V_{\rm sat}$, therefore, the apparent 
magnitude of satellites, $m_{\rm sat}$, is written as follows: 

\begin{eqnarray}
m_{\rm sat} = m_{\rm star} - 2.5 \log\biggr(\frac{V_{\rm sat}}{V_{\rm star}}\frac{f_{\rm sat}}{f_{\rm star}}\biggr), 
\end{eqnarray}
where $V_{\rm star}, f_{\rm star}$, and $m_{\rm star}$ are the angular velocity along 
the celestial sphere, flux, and magnitude of reference stars (i.e., $V_{\rm star}=
15\times\cos\delta~{\rm arcsec~s^{-1}},~\delta:$ declination), respectively. Here, we observed 
the satellites with star tracking, and after that, the reference stars were observed by stopping 
the star tracking immediately. The velocity of the reference stars during fixed observation is 
$V_{\rm star}$. It is possible to estimate the angular velocity, $V_{\rm sat}$, by calculating a traverse 
speed on the great-circular distance, $\lambda$, which is written as follows: 
\begin{eqnarray}
\lambda = \arccos(\sin\delta \sin D + \cos\delta\cos D\cos(A-\alpha)), 
\end{eqnarray}
where ($\alpha,~\delta$) and ($A,~D$) are the right ascension, declination of satellites at 
the initial time of an observation and those at a certain time, respectively. In order to 
measure the flux of reference stars, $f_{\rm star}$, by the same way to evaluate flux of 
satellite trails, we elongated the CCD images of reference stars so that the images become 
the same shape as satellite light trails. Figures 1, 2, 3, and 4 show the FITS images of 
Darksat and elongated reference stars around the satellite trail. In Figures 1 and 4, we do 
not include the bright stars on a right-side edge of images for reference stars due to a count bias 
on the CCD image of the $I_{\rm c}$ band. Since there are no bright reference stars without the 
count bias around the STARLINK-1113 trail (Figure 4), we added and observed a bright reference 
star near the target field (i.e., three reference stars for the STARLINK-1113 trail, see Table 2). 

\subsection{Data Analysis}
We performed dark subtraction, dome flat corrections and sky subtraction with Image 
Reduction and Analysis Facility (IRAF). In the actual data analysis, we subtracted    
bad pixels and stars from the original satellite trail images (see middle panels 
of Figures 1, 2, 3, and 4). In this study we adopted the star catalog UCAC4 that is, 
for instance, able to refer to Johnson's $B$, $V$, SDSS $r$, and $i$ band magnitudes.
It is possible to derive $g'$-, $R_{\rm c}$- and $I_{\rm c}$-band magnitudes using the 
equations in Table 3 of \citet{2006A&A...460..339J}:
\begin{eqnarray}
g = V + 0.630(B-V) - 0.124, \\
R =  r - 0.252(r-i) - 0.152, \\
I = r  - 1.245(r-i) - 0.387. 
\end{eqnarray}
We measured the count profile perpendicular to the streak for each object, 
and compared the corresponding profile from the elongated star image. 
We note again that the flux per unit length of the streak is inversely proportional 
to the angular velocity of an object (see Equation (1)). We used $\tt{Projection}$ in 
SAOImage~DS9\footnote{https://sites.google.com/cfa.harvard.edu/saoimageds9} 
to measure the flux (averaged counts of cross sections) in a 
rectangular-region for the integration along the Darksat and STARLINK-1113 trails, 
and the reference stars around these satellite trails. The rectangular region to measure the 
averaged satellite flux, $f_{\rm sat}$, is 10 times thicker than that of the 
reference stars but avoiding the edge of CCDs (i.e., the trail length was limited in measuring  
its flux), and then we integrated the area under the profile. Figure 5 exhibits averaged 
CCD counts along the Darksat and STARLINK-1113 trails. 

The statistical magnitude errors, $\sigma_{m}$, of Darksat and STARLINK-1113 were 
estimated by applying the law of error propagation to Equation (1): 
\begin{eqnarray}
\sigma_{m} = \frac{2.5}{\ln10}\sqrt{\biggr(\frac{\delta f_{\rm sat}}{f_{\rm sat}}\biggr)^2+
\biggr(\frac{\delta f_{\rm star}}{f_{\rm star}}\biggr)^2+\biggr(\frac{\delta V_{\rm sat}}{V_{\rm sat}}\biggr)^2}, 
\end{eqnarray}
where $\delta f_{\rm sat}, \delta f_{\rm star}$, and $\delta V_{\rm sat}$ are the flux errors of 
light trails of satellites and reference stars, and the velocity error of satellites, respectively.
In this study, these errors take the following ranges: 0.005 $\lesssim \delta f_{\rm sat}/f_{\rm sat}
\lesssim 0.07,~0.004\lesssim\delta f_{\rm star}/f_{\rm star}\lesssim 0.21,$ and $0.002 \lesssim 
\delta V_{\rm sat}/V_{\rm sat} \lesssim 0.06$. Since the angular velocity of the reference stars, 
$V_{\rm star}$, is constant, we ignored in Equation (6) the term related with the angular velocity of 
stars. We estimated the standard deviation of sky flux, $\sigma_{\rm sky}$, around the satellite trails 
to evaluate the flux errors of the trails, $\delta f_{\rm sat}$, and those of reference stars, 
$\delta f_{\rm star}$. Here, the sky region to estimate $\sigma_{\rm sky}$ has the same rectangular 
shape when we estimated the flux of trails by satellites or reference stars (i.e., $\delta f_{\rm sat}$ or 
$\delta f_{\rm star} = \sigma_{\rm sky} \times {\rm pixel~width~of~trails}$). The velocity error of 
satellites, $\delta V_{\rm sat}$, is an averaged value of the velocity difference between the central to  
start exposure time, $|V_{\rm sat}-V_{\rm start}|$, and that of the central to end time, 
$|V_{\rm sat}-V_{\rm end}|$. 

\begin{figure*}
\figurenum{1}
\begin{flushleft}
 \includegraphics[height=15cm,width=18cm]{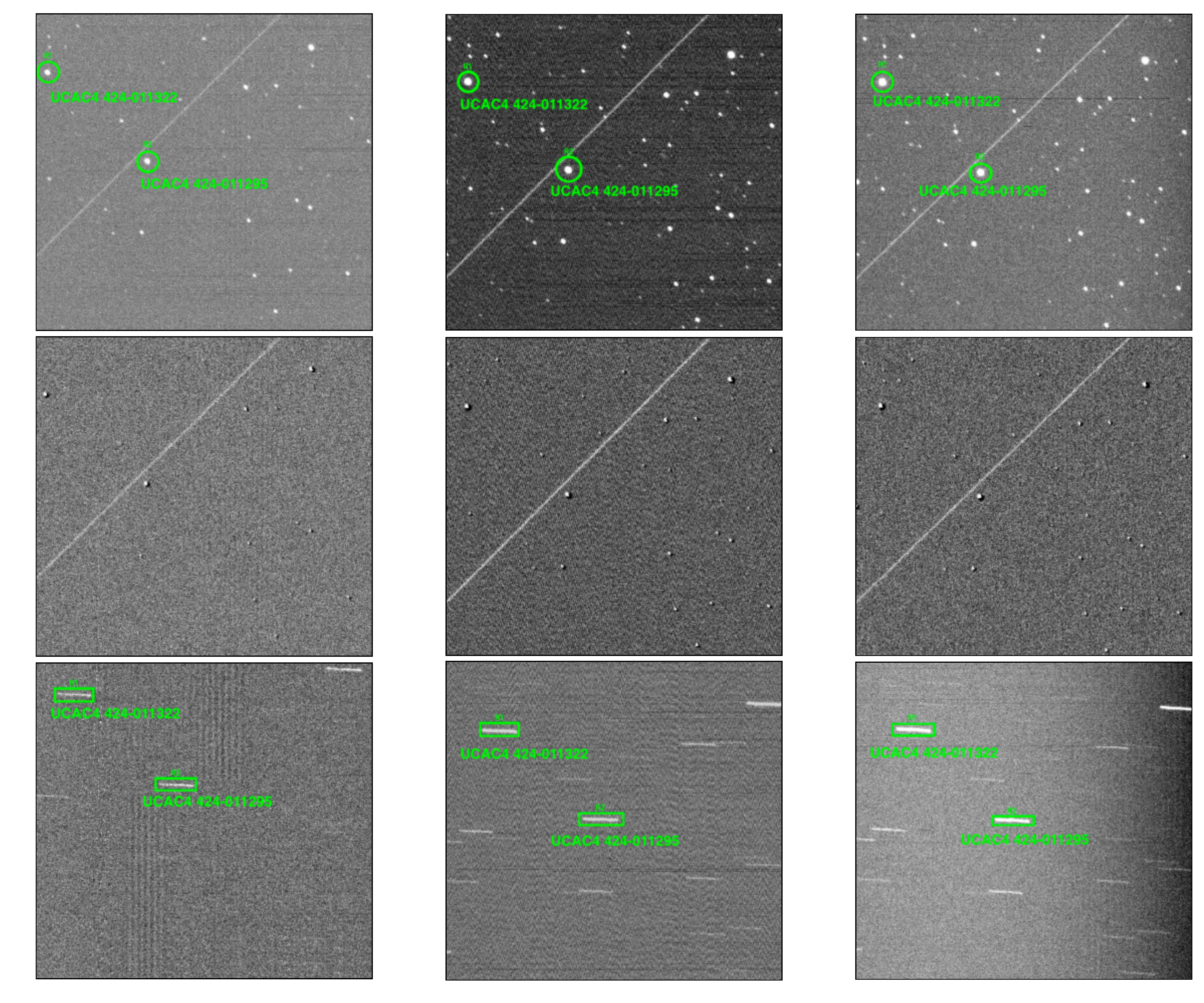} 
\caption{The FITS images of Darksat (top), bad-pixel-subtracted images (middle), 
and elongated reference stars (bottom) in April 10, 2020 taken by the Murikabushi 
telescope/$\it{MITSuME}$ $g'$ (left), $R_{\rm c}$ (center), and $I_{\rm c}$ (right) 
bands. Reference stars around the satellite trails are marked with circles (top) or 
rectangles (bottom). 
\label{fig:f1}}
\end{flushleft}
\end{figure*}
 
\begin{figure*}
\figurenum{2}
\begin{flushleft}
 \includegraphics[height=15cm,width=18cm]{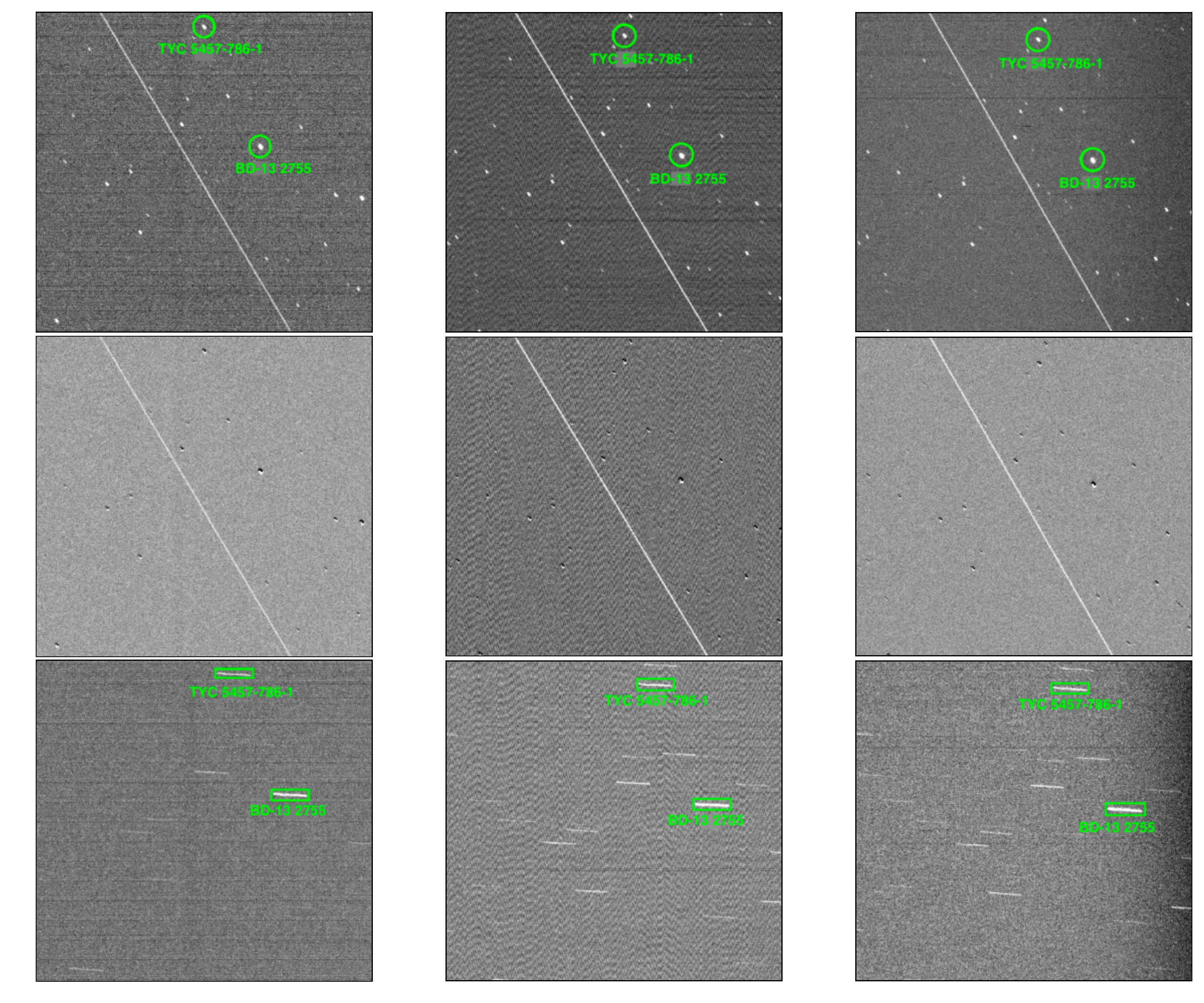} 
\caption{Same as Figure 1, but for the Darksat images in May 18, 2020. 
\label{fig:f1}}
 \end{flushleft}
 \end{figure*}
 
 \begin{figure*}
\figurenum{3}
\begin{flushleft}
 \includegraphics[height=15cm,width=18cm]{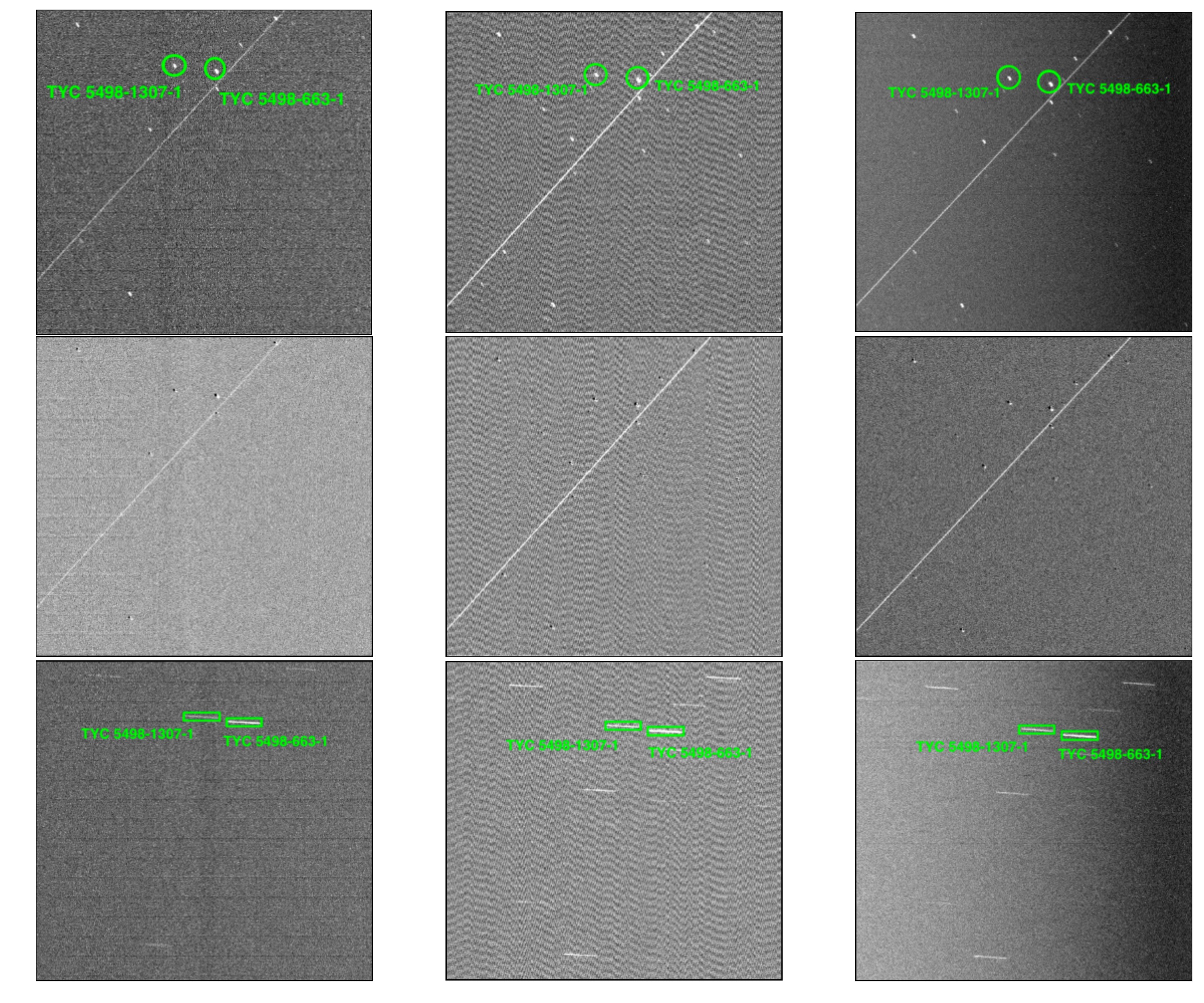} 
\caption{Same as Figure 1, but for the Darksat images in June 10, 2020. 
\label{fig:f1}}
 \end{flushleft}
 \end{figure*}
 
 \begin{figure*}
\figurenum{4}
\begin{flushleft}
 \includegraphics[height=15cm,width=18cm]{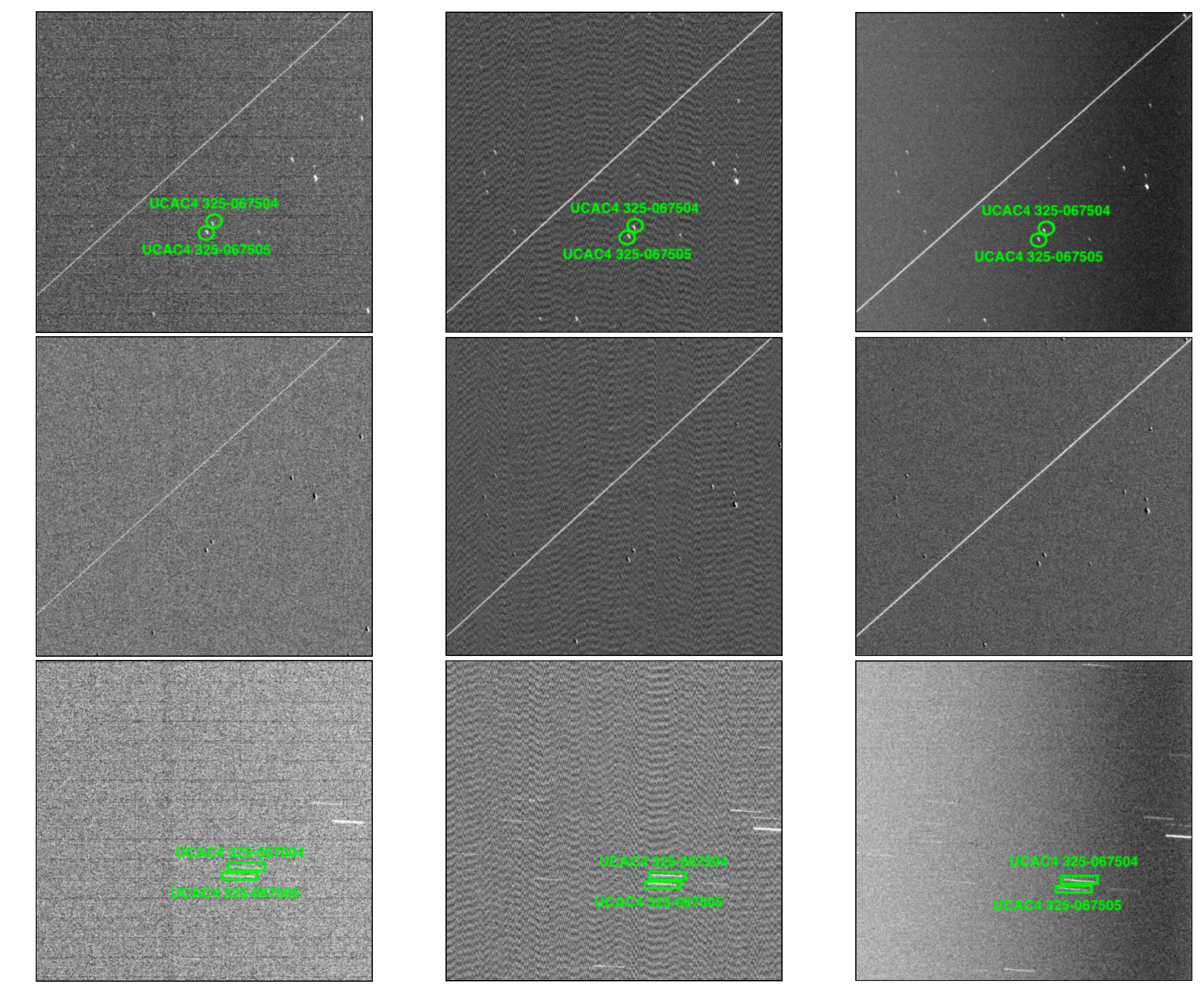} 
\caption{Same as Figure 1, but for the STARLINK-1113 images in June 10, 2020. 
\label{fig:f1}}
 \end{flushleft}
 \end{figure*}
 
\begin{figure*}
\figurenum{5}
\begin{flushleft}
 \includegraphics[height=15cm,width=18cm]{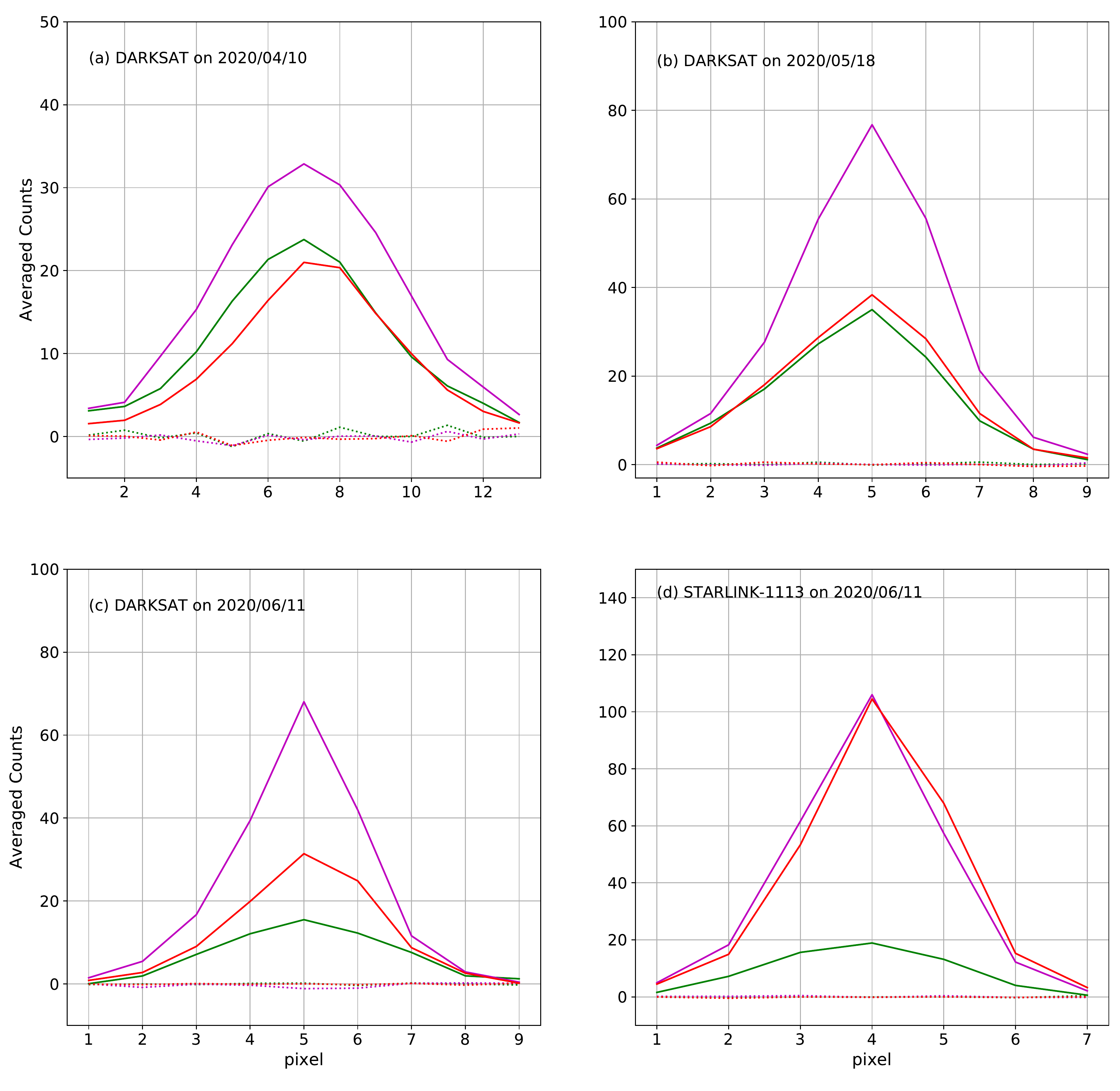} 
\caption{Averaged-section counts along the Darksat and STARLINK-1113 trails in the Murikabushi 
telescope/$\it{MITSuME}$ $g'$ (green), $R_{\rm c}$ (magenta), and $I_{\rm c}$ (red) bands. 
Solid and dotted lines indicate counts of the satellite trails and sky region around the trails, 
respectively.
\label{fig:f1}}
 \end{flushleft}
 \end{figure*}

\begin{deluxetable*}{cccccc}[b!]
\tablecaption{Observation log of Darksat and STARLINK-1113 by the Murikabushi 
telescope/$\it{MITSuME}$ $g', R_{\rm c}$ and $I_{\rm c}$ bands \label{tab:mathmode}}
\tablecolumns{5}
\tablenum{1}
\tablewidth{0pt}
\tablehead{
\colhead{}&
\colhead{Darksat}&
\colhead{Darksat}&
\colhead{Darksat}&
\colhead{STARLINK-1113}&
%\colhead{} & \colhead{(d)} &
%\colhead{(arcsec)} & \colhead{} & \colhead{} 
}
\startdata
Observation Date & April 10, 2020& May 18, 2020 & June 11, 2020 & June 11, 2020 \\ 
Start Time of Observation (UTC) & 10:53:50 & 11:13:55 & 12:31:56 & 12:17:56\\ 
Central Time of Observation (UTC) & 10:54:00 & 11:14:00 & 12:31:58.5 & 12:17:58.5 \\ 
End Time of Observation (UTC) & 10:54:10 & 11:14:05 & 12:32:01 &12:18:01 \\ 
Exposure Time (s) & 20.0 & 10.0 & 5.0 & 5.0 \\
RA$^a$& 05:52:35.05 & 09:04:46.3 & 10:42:36.7 & 12:23:17.5 \\ 
Dec$^a$ & -05:18:43.8 & -13:58:57.0 & -13:34:37.2 & -25:03:20.0 \\  
Azimuth$^{a,~b}$ ($^\circ$) & 236.67 & 223.29 & 238.57 & 205.66 \\ 
Elevation$^a$ ($^\circ$) & 42.49 & 40.00 & 26.96  & 35.80 \\ 
Airmass$^a$ & 1.48 & 1.55 & 2.20 & 1.71 \\ 
Altitude$^a$ (km) & 550.16 & 549.88 & 549.72 & 549.56 \\
Distance between Satellite and Observer (km) & 781.54~$\pm$~3.12 & 812.03~$\pm$~25.62 
& 1068.69~$\pm$~0.91 & 873.41~$\pm$~6.36 \\ 
Solar Phase Angle$^a$ [Sun-Target-Observer] ($^\circ$) & 109.9 & 93.7 & 93.5 & 67.4 \\
Phase Angle$^a$ [Sun-Observer-Target] ($^\circ$) & 70.1 & 86.3 & 86.5  & 112.6 \\
Observer Phase Angle$^a$ ($^\circ$) & 42.9 & 44.8 & 55.1 & 48.3 \\  
Angular Velocity (arcsec s$^{-1}$) & 1934.10~$\pm$~10.88 & 1322.99~$\pm$~76.20 
& 1415.33~$\pm$~2.35 & 1626~$\pm$~22.03 \\
\hline 
\enddata
\tablenotetext{a}{Values at the central time of observations.}
\tablenotetext{b}{The north and the east are 0$^\circ$ and 90$^\circ$, respectively.}
\end{deluxetable*}

\section{Results}
In this section, we provide the multi-band magnitudes and colors of the Darksat 
and STARLINK-1113 trails. 

\subsection{Apparent and Normalized Magnitude}
First, we measured the apparent magnitudes of Darksat and STARLINK-1113 
as described in the previous section. In addition to correcting the difference 
of the distance between an observer and a satellite, we normalized these 
magnitudes at the satellite orbital altitude of $\sim$550 km (hereafter, normalized 
magnitude) by adding a factor of +5$\log(r/550)$, where $r$ is the distance between 
the satellite and observer \citep[see also Tr20 and][]{2020arXiv200612417T}. 
We considered the errors of $r$ for the estimation of the normalized magnitudes. 
Table 2 summarizes the reference stars, apparent, and normalized $g'$-, 
$R_{\rm c}$-, and $I_{\rm c}$-band magnitudes for each observation epoch. The 
apparent and normalized $g'$-band magnitudes of Darksat ($7.37~\pm~0.10$ and 
$6.35~\pm~0.10$ mag on average) are slightly brighter than that of STARLINK-1113 
($7.69~\pm~0.16$ and $6.68~\pm~0.17$ mag) in our measurements, while those 
of Darksat are fainter than STARLINK-1113 in $R_{\rm c}$ and $I_{\rm c}$ 
bands. We found a tendency that the longer the observed wavelength is, the brighter the 
two satellite magnitudes become. Especially, STARLINK-1113 showed the extremely 
bright $I_{\rm c}$-band magnitude of 5.25~$\pm$~0.07 (apparent magnitude) and 
4.25~$\pm$~0.07 mag (normalized magnitude). 

Next, we estimated the colors of the Darksat and 
STARLINK-1113 trails. Under the simultaneous multicolor observations, the largest 
advantage of color estimations is that the effects on each band can be 
tested without considering the dependence on the all parameters such as a satellite orbital 
altitude (or the distance $r$), angular velocity, and solar (or observer) phase angle. 
Table 3 lists the colors $g'-R_{\rm c}$, $g'-I_{\rm c}$, and $R_{\rm c}-I_{\rm c}$ of Darksat 
and STARLINK-1113 in each observation epoch. In the case of Darksat, while there is no 
significant color difference between May 18 and June 11, that of April 10 is different from the 
data of the other two epochs due to the weather condition. The colors $g'-I_{\rm c}$, and 
$R_{\rm c}-I_{\rm c}$ of STARLINK-1113 are extremely ($>$ 1 mag) redder than that of 
Darksat for any observation epochs. 

\subsection{Phase Angle Effect}
We used the orbital information in HORIZONS 
Web-Interface\footnote{https://ssd.jpl.nasa.gov/horizons.cgi} in order to examine solar 
phase angle dependence of the normalized magnitudes (see also Table 1). In Figure 6 we 
plotted our data point on the solar phase angles vs normalized $g'$-, $R_{\rm c}$-, and 
$I_{\rm c}$-band magnitudes plane together with the results of Tr20. The $g'$, and 
$R_{\rm c}$-band magnitudes of Darksat do not show any clear relation with the solar 
phase angle. Those of $g'$ band are consistent with that of Tr20 despite a solar phase angle 
difference of $\sim 40^\circ - 55^\circ$. On the other hand, the $g'$-band magnitude of 
STARLINK-1113 is $\sim$ 1.5 mag darker than that of Tr20 in spite of the smaller one 
of $\sim 6^\circ$. Consequently, the multi-band magnitudes do not exhibit a clear 
correlation with solar phase angles. The phase angle effect alone cannot explain the 
magnitude difference between Darksat and STARLINK-1113 (see Appendix A about both the 
the solar and observer phase-angle effects). This may be due to the small 
number of observation points, and it would be needed to conduct more frequent 
observations to reveal if there is a relation between satellite brightness and solar phase angle. 

\begin{deluxetable*}{cccccccc}[b!]
\tablecaption{Magnitudes of Darksat, STARLINK-1113, and reference stars \label{tab:mathmode}}
\tablecolumns{7}
\tablenum{2}
\tablewidth{0pt}
\tablehead{
\colhead{Reference Star}&
\colhead{$g'$ magnitude$^a$}&
\colhead{$g'$ magnitude}&
\colhead{$R_{\rm c}$ magnitude$^a$}&
\colhead{$R_{\rm c}$ magnitude}&
\colhead{$I_{\rm c}$ magnitude$^a$}&
\colhead{$I_{\rm c}$ magnitude}&\\
\colhead{}& 
\colhead{(Star)}&
\colhead{(Satellite)}&
\colhead{(Star)}&
\colhead{(Satellite)}&
\colhead{(Star)}&
\colhead{(Satellite)}&
%\colhead{} & \colhead{(d)} &
%\colhead{(arcsec)} & \colhead{} & \colhead{} 
}
\startdata
\multicolumn{7}{c}{Date: April 10, 2020, Target: Darksat} \\ \hline 
UCAC4 424-011322 & 12.53 & 6.98~$\pm$~0.11 & 10.68  & 6.92~$\pm$~0.04 & 9.75 & 6.62~$\pm$~0.07 \\ 
UCAC4 424-011295 & 12.41 & 6.91~$\pm$~0.11 & 10.93 & 6.91~$\pm$~0.04 & 10.16 & 6.65~$\pm$~0.07 \\  \hline 
Apparent Magnitude$^b$& $\ldots$ & 6.95~$\pm$~0.11 & $\ldots$ & 6.92~$\pm$~0.04 & $\ldots$ & 6.64~$\pm$~0.07 \\  
Normalized Magnitude& $\ldots$ & 6.18~$\pm$~0.11 & $\ldots$ & 6.15~$\pm$~0.04 & $\ldots$ & 5.87~$\pm$~0.07 \\  \hline 
\multicolumn{7}{c}{Date: May 18, 2020, Target: Darksat} \\ \hline
BD-13 2755 & 10.89 & 7.66~$\pm$~0.07 & 10.05 & 7.03~$\pm$~0.06 & 9.59 & 6.58~$\pm$~0.07 \\ 
TYC 5457-786-1 & 12.32 & 7.39~$\pm$~0.07 & 11.77 & 6.96~$\pm$~0.06 & 11.38 & 6.59~$\pm$~0.08 \\  \hline  
Apparent Magnitude$^b$ & $\ldots$ & 7.52~$\pm$~0.07 & $\ldots$ & 6.99~$\pm$~0.06 & $\ldots$ & 6.59~$\pm$~0.07 \\  
Normalized Magnitude &  $\ldots$ & 6.68~$\pm$~0.10 & $\ldots$ & 6.15~$\pm$~0.09 & $\ldots$ & 5.74~$\pm$~0.10 \\  \hline  
\multicolumn{7}{c}{Date: June 11, 2020, Target: Darksat} \\ \hline
TYC 5498-663-1 & 11.33 & 7.76~$\pm$~0.04 & 10.77 & 7.15~$\pm$~0.03 & 10.45 & 6.71~$\pm$~0.02 \\ 
TYC 5498-1307-1 & 12.54  & 7.53~$\pm$~0.16 & 12.03 & 7.12~$\pm$~0.04 & 11.69 & 6.83~$\pm$~0.05 \\  \hline 
Apparent Magnitude$^b$ & $\ldots$ & 7.65~$\pm$~0.12 & $\ldots$ & 7.13~$\pm$~0.04 & $\ldots$ & 6.77~$\pm$~0.04 \\  
Normalized Magnitude& $\ldots$ & 6.20~$\pm$~0.12 & $\ldots$ & 5.69~$\pm$~0.04 & $\ldots$ & 5.33~$\pm$~0.04 \\  \hline 
\multicolumn{7}{c}{Average value of the three epochs} \\ \hline
Apparent Magnitude$^b$ & $\ldots$ & 7.37~$\pm$~0.10 & $\ldots$ & 7.01~$\pm$~0.05 & $\ldots$ & 6.66~$\pm$~0.06 \\  
Normalized Magnitude& $\ldots$ & 6.35~$\pm$~0.10 & $\ldots$ & 6.00~$\pm$~0.06 & $\ldots$ & 5.65~$\pm$~0.08 \\  \hline 
\multicolumn{7}{c}{Date: June 11, 2020, Target: STARLINK-1113} \\ \hline
UCAC4 325-067504 & 13.76 & 7.59~$\pm$~0.23 & 12.44 & 6.59~$\pm$~0.05 & 11.72 & 5.12~$\pm$~0.07 \\ 
UCAC4 325-067505 &  13.18 & 7.48~$\pm$~0.16 & 12.61 & 6.62~$\pm$~0.10 & 12.17 & 5.10~$\pm$~0.10 \\  
CD-24 10321 & 10.42 & 8.00~$\pm$~0.04 & 9.37 & 6.62~$\pm$~0.02 & 8.66 & 5.53~$\pm$~0.02 \\ \hline 
Apparent Magnitude$^b$ & $\ldots$ & 7.69~$\pm$~0.16 & $\ldots$ & 6.61~$\pm$~0.06 & $\ldots$ & 5.25~$\pm$~0.07 \\  
Normalized Magnitude& $\ldots$ & 6.68~$\pm$~0.17 & $\ldots$ & 5.60~$\pm$~0.06 & $\ldots$ & 4.25~$\pm$~0.07 \\  \hline  
\enddata
\tablenotetext{a}{Converted UCAC4 magnitudes by Equations (3), (4), and (5).}
\tablenotetext{b}{Averaged magnitude with the above reference stars.}
\end{deluxetable*}

\begin{deluxetable*}{cccccc}[b!]
\tablecaption{Colors of Darksat and STARLINK-1113 \label{tab:mathmode}}
\tablecolumns{5}
\tablenum{3}
\tablewidth{0pt}
\tablehead{
\colhead{Target Satellite}&
\colhead{Date}&
\colhead{$g'-R_{\rm c}$}&
\colhead{$g'-I_{\rm c}$}&
\colhead{$R_{\rm c}-I_{\rm c}$}
}
\startdata
Darksat & April 10, 2020 &0.03~$\pm$~0.12 & 0.31~$\pm$~0.13 & 0.28~$\pm$~0.08 \\  
Darksat & May 18, 2020 & 0.53~$\pm$~0.09 & 0.93~$\pm$~0.10 & 0.41~$\pm$~0.10 \\    
Darksat & June 11, 2020 & 0.51~$\pm$~0.13 & 0.87~$\pm$~0.13 & 0.36~$\pm$~0.05 \\  \hline 
Average & $\ldots$ &  0.36~$\pm$~0.11 & 0.71~$\pm$~0.12  & 0.35~$\pm$~0.08  \\  \hline 
STARLINK-1113 & June 11, 2020 & 1.08~$\pm$~0.18 & 2.44~$\pm$~0.18 & 1.36~$\pm$~0.09 \\ \hline 
\enddata
\end{deluxetable*}

\begin{figure*}
\figurenum{6}
\begin{flushleft}
 \includegraphics[height=12cm,width=18cm]{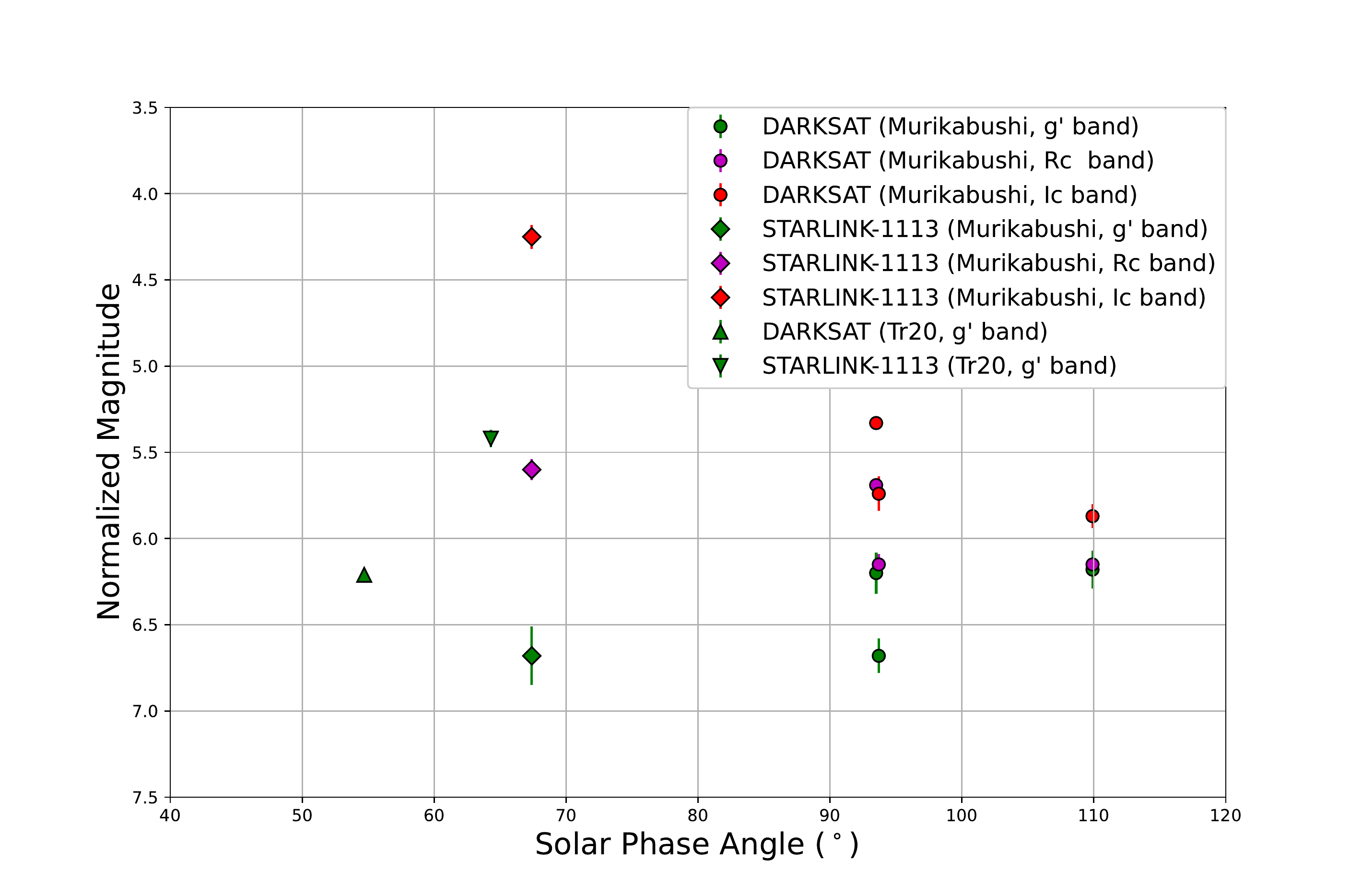} 
\caption{A plot between the solar phase angle and normalized $g'~(green)$-, 
$R_{\rm c}~(magenta)$-, and $I_{\rm c}~(red)$-band magnitudes. 
\label{fig:f1}}
 \end{flushleft}
 \end{figure*}
 
\section{Discussion}

\subsection{The effect of Atmospheric Extinction} 
The STARLINK-1113 trail showed extremely redder color than the Darksat trail (Table 3). 
We will discuss below the effect of atmospheric extinction. The STARLINK-1113 trail was 
observed on June 11, 2020 12:18:00 UTC in this study. Shortly after 12:18:10 UTC, this 
satellite plunged into the Earth shadow. As shown in Figure 7, we define the following 
angles: (1) angle, $\theta_1$, between the line from the geocenter to ground surface 
(i.e., the Earth radius $R_\oplus=6371$ km) and the line from the geocenter to the satellite 
orbital altitude at the beginning of the Earth shadow $l$ of 6921 km (point B), (2) angle, 
$\theta_2$, between the line from the geocenter to the upper edge of the atmosphere 
(i.e., $R_\oplus+\Delta R$ of 6471 km) and the line from the geocenter to the intersecting 
point with the tangent line of the upper atmosphere (point A; the same length as $l$). 
A range of angle, $\theta_1-\theta_2$, where atmospheric extinction becomes noticeable 
is written as follows: 
\begin{eqnarray}
\theta_1-\theta_2 = \arctan{\biggr(\frac{\sqrt{l^2-R_\oplus^2}}{R_\oplus}\biggr)} - \nonumber\\
\arctan{\biggr(\frac{\sqrt{l^2-(R_\oplus+\Delta R)^2}}{R_\oplus +\Delta R}\biggr)} \sim 2.22~[{\rm deg}].
\end{eqnarray}\\
Since the orbital period of STARLINK-1113 was 95.64 min at the observation time, 
the angular velocity of STARLINK-1113 was $\sim$ 0.06 deg s$^{-1}$; the time 
required to cross the range, $\theta_1-\theta_2$, is $\sim$35.4 s. 
Namely, STARLINK-1113 was orbiting in the vicinity of the entrance to the Earth shadow 
at our observation time, and was strongly affected by the atmospheric scattering in the 
$g'$, and $R_{\rm c}$ bands. The Earth shine and terrestrial radiation would 
also contribute magnitudes of satellite trails. However, the contribution of these radiation 
can be reduced in the Earth shadow. Other than the altitudes and the phase angles of 
satellite, the effect of atmospheric extinction should also be taken into account as 
another factor. In other words, the brightness of the usual Starlink's LEO 
satellites can be reduced comparable to that of Darksat by taking into account the 
above effects. 

 \begin{figure*}
\figurenum{7}
\begin{center}
 \includegraphics[height=9cm,width=16cm]{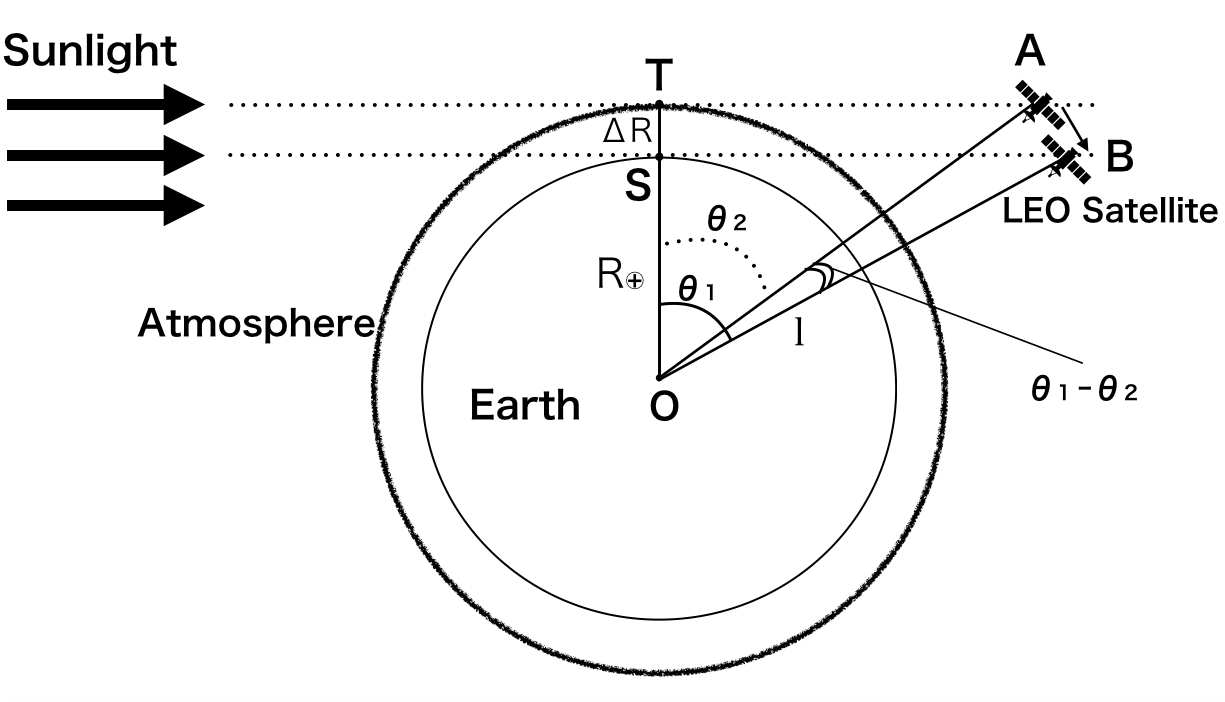} 
\caption{The LEO satellite plunging into the Earth shadow (i.e., satellite from point A 
to B). Points O, S, and T are the geocenter, ground surface, and Kármán line, 
respectively. The following scales and angles are defined in this figure: the Earth radius,  
$R_\oplus =|{\rm OS}|=$ 6,371 km, atmospheric thickness, $\Delta 
R=|{\rm ST}|=100$ km, orbital height from the geocenter, $l=|\rm OA|=
|\rm OB|=$ 6,921 km, $\theta_1$ ($\angle$BOS), and $\theta_2$ ($\angle$AOT). 
The LEO satellite orbits the region formed by $\angle$BOA ($\theta_1-\theta_2$) 
where the Sunlight to the satellite is strongly affected by the Earth's atmosphere.
\label{fig:f1}}
\end{center}
\end{figure*} 

\subsection{Modeling the Flux of Darksat and STARLINK-1113}
\subsubsection{Effective Radiation Temperature of Darksat and STARLINK-1113}

%%%%%%%%
We evaluate the surface temperature of Darksat and STARLINK-1113 from the 
apparent magnitudes, $m_{\rm Sat}$, in Table 2. Although these LEO satellites 
have a shape of flat panel, if we assume the satellites are spherical, the surface 
temperature, $T_{\rm sat}$, is written as follows \citep{1989Icar...78..337S}: 
\begin{eqnarray}
T_{\rm sat} = \biggr(\frac{I(1-a_{\rm obs})}{4\sigma\epsilon}\biggr)^{\frac{1}{4}}, 
\end{eqnarray}
where $I~(=1.37\times10^{3}$~W~m$^{-2}$), $a_{\rm obs}$, $\epsilon~(=0.9)$, and 
$\sigma$ ($=5.67\times10^{-8}$~W m$^{-2}$ K$^{-4}$) are the solar constant, measured 
surface albedo, the infrared emissivity \citep{1986Icar...68..239L}, and the 
Stefan-Boltzmann constant, respectively. For estimating of the surface temperature, 
$T_{\rm sat}$, we need to derive the absolute magnitude of these satellites, $H$: 
\begin{eqnarray}
H = m_{\rm Sun} - 2.5 \log\frac{a_{\rm obs}~{r_{\rm sat}}^2}{(1 {\rm au})^2},
\end{eqnarray}
where $m_{\rm Sun}$ and $r_{\rm sat}$ are the apparent magnitude of the 
Sun \citep{2018ApJS..236...47W} and the radius of Darksat or 
STARLINK-1113, respectively. 
We employed the radius, $r_{\rm sat}$, of 1.5 m for the Starlink satellites  
\citep[i.e., the flat panel with a diameter of 3 m; ][]{2020ApJ...892L..36M}. 
The $H$ for the LEO satellites is also approximately described by the phase integral, 
$p(\theta)$, as a first-approximation function of the solar phase angle $\theta$ \citep{1907Obs....30...96W}:  
\begin{eqnarray}
m_{\rm sat} \sim  H - 2.5 \log \biggr(\frac{(1~{\rm au})^2~p(\theta)}{r^2}\biggr), \\
\nonumber\\
p(\theta) = \frac{2}{3} \biggr(\biggr(1-\frac{\theta}{\pi}\biggr)\cos\theta + \frac{1}{\pi}\sin\theta \biggr), 
\end{eqnarray}
where $r$ is the distance between an observer and a satellite. It is able to derive the 
surface albedo, $a_{\rm obs}$, from Equations (9) through (11). Then, the surface 
temperature, $T_{\rm sat}$, is evaluated immediately. Table 4 summarizes the surface 
temperature of Darksat and STARLINK-1113 in this study. From Table 4, we adopt the 
temperature, $T_{\rm sat}$, of 280 K to be used in the following discussion. 
We note the satellite radius of 1.5 m is an approximate value and has an uncertainty, 
$\delta r_{\rm sat}$. From Equations (8) through (10), the surface temperature is inversely 
proportional to the square root of the satellite radius ($T_{\rm sat} \propto r_{\rm sat}^{-1/2}$). 
When assuming the uncertainty $\delta r_{\rm sat}$ of 0.5 m, corresponding uncertainty  
of the surface temperature will be at least  2 $\sim$ 3 degrees, which is comparable to the 
standard deviation of the surface temperature listed in 
Table 4 ($\sigma_{T_{\rm sat}}\sim2$~K). \\
 
\begin{deluxetable*}{cccccc}[b!]
\tablecaption{The phase integral, absolute magnitude, measured albedo, and surface 
temperature of Darksat and STARLINK-1113 \label{tab:mathmode}}
\tablecolumns{5}
\tablenum{4}
\tablewidth{0pt}
\tablehead{
\colhead{Target Satellite (Filter, Date)}&
\colhead{Phase Integral $p(\theta)$}&
\colhead{Absolute Magnitude $H$$^a$}&
\colhead{Albedo $a_{\rm obs}^a$}&
\colhead{Temperature $T_{\rm sat}^a$}\\
\colhead{}&
\colhead{}&
\colhead{}&
\colhead{}&
\colhead{(K)} 
}
\startdata
Darksat ($g'$-band, April 10, 2020) & 0.11 & 30.97 & 0.12 & 277.38 \\  
Darksat ($g'$-band, May 18, 2020) & 0.19 & 32.05 & 0.04 & 283.04 \\    
Darksat ($g'$-band, June 11, 2020) & 0.19 & 31.59 & 0.07 & 281.31 \\  \hline 
Darksat ($R_{\rm c}$-band, April 10, 2020) & 0.11 & 30.94 & 0.06 & 282.01 \\  
Darksat ($R_{\rm c}$-band, May 18, 2020) & 0.19 & 31.52 & 0.03 & 283.77 \\    
Darksat ($R_{\rm c}$-band, June 11, 2020) & 0.19 & 31.07 & 0.05 & 282.49 \\  \hline 
Darksat ($I_{\rm c}$-band, April 10, 2020) & 0.11 & 30.66 & 0.06 & 282.17 \\  
Darksat ($I_{\rm c}$-band, May 18, 2020) & 0.19 & 31.12 & 0.04 & 283.58  \\    
Darksat ($I_{\rm c}$-band, June 11, 2020) & 0.19 & 31.71 & 0.05 & 282.34 \\  \hline 
Averaged Temperature & $\ldots$ & $\ldots$ & $\ldots$ & 282.01 \\  \hline 
STARLINK-1113 ($g'$-band, June 11, 2020) & 0.36 & 32.73 & 0.02 & 284.55 \\  
STARLINK-1113 ($R_{\rm c}$-band, June 11, 2020) & 0.36 & 31.66 & 0.03 & 284.06 \\    
STARLINK-1113 ($I_{\rm c}$-band, June 11, 2020) & 0.36 & 30.30 & 0.08 & 280.49 \\  \hline 
Averaged Temperature & $\ldots$ & $\ldots$ & $\ldots$ & 283.07 \\  \hline 
\enddata
\tablenotetext{a}{The radius of the satellites, $r_{\rm sat}$, is fixed to 1.5 m.}
\end{deluxetable*}

%%%%%%%%%%%

\subsubsection{Flux model by The Blackbody Radiation}
In this section, we model the normalized $g'$-, $R_{\rm c}$, and $I_{\rm c}$-band 
flux of the Darksat and STARLINK-1113 trails in each epoch with the blackbody radiation. 
We derived these flux from the normalized and AB magnitudes in each band 
\citep{2007AJ...133..734B}. When modeling the flux of the satellite trails with 
blackbody radiation, the following four components were  
considered: thermal radiation of the satellite, $F_{\rm TS}$, reflection of the sunlight,  
$F_{\rm RS}$, Earthshine, $F_{\rm REs}$, and reflection of Earth thermal radiation, $F_{\rm TE}$. 
The four components are written as follows \citep[see also][]{2003A&A...397..325A}: 

{\footnotesize
\begin{eqnarray}
F_{\rm TS} = \pi\epsilon\biggr(\frac{r_{\rm sat}}{h_{\rm T}}\biggr)^2B(\lambda,~T_{\rm sat})~\frac{\lambda^2}{c}, \\
F_{\rm RS} = \pi\biggr(\frac{R_{\odot}}{1~{\rm au}}\biggr)^2B(\lambda,~T_{\odot})a_{\rm mod}
~p(\theta)\biggr(\frac{r_{\rm sat}}{h_{\rm T}}\biggr)^2~\frac{\lambda^2}{c},\\ 
F_{\rm REs} = a_{\rm E}\biggr(\frac{R_{\oplus}}{R_{\oplus}+h_{\rm T}}\biggr)^2\biggr\{1-\biggr(\frac{R_{\oplus}}{R_{\oplus}+h_{\rm T}}\biggr)^2\biggr\}~
\frac{p(\chi)}{p(\theta)}~F_{\rm RS},\\
F_{\rm TE} = \pi\epsilon\biggr(\frac{R_{\oplus}}{R_{\oplus}+h_{\rm T}}\biggr)^2B(\lambda,~T_{\rm E})
~a_{\rm mod}\biggr(\frac{r_{\rm sat}}{h_{\rm T}}\biggr)^2\frac{\lambda^2}{c},
\end{eqnarray}
\normalsize}

where $T_{\odot}~(=5772~{\rm K})$, $T_{\rm E}~(=290~{\rm K})$, $R_{\odot}~(=7.0\times10^5~{\rm km})$, 
$R_{\oplus}$, $h_{\rm T}~(=550~{\rm km})$, $a_{\rm mod}$, $a_{\rm E}~(=0.3)$, and $p(\chi)$ are 
the temperature of the Sun and the Earth, radius of the Sun and the Earth, orbital height of the LEO satellites, 
modeled albedo with the blackbody, albedo of the Earth, and phase integral as a function of the 
Sun-observer-target phase angle $\chi$, respectively. The blackbody radiation $B(\lambda,~T)$ 
is expressed by:
\begin{eqnarray}
B(\lambda,~T) = \frac{2hc^2}{\lambda^5}\frac{1}{\exp{(\frac{hc}{kT\lambda})}-1},
\end{eqnarray}
where $c$, $\lambda$, $h$, and $k$ are the speed of light, wavelength, the Planck constant, 
and the Boltzmann constant, respectively. We note that all of the four components (i) are  
monochromatic flux density per unit frequency, and (ii) are multiplied by a factor, $(r/h_{\rm T})^2$, 
for normalization with the orbital height, $h_{\rm T}$. For the derivation of 
Equations (14) and (15), see Appendix B. The Moon-light effect can be ignored 
for all three observation epochs, since the Moon was below the horizon at the observation 
times (see Appendix C about the Moon effects). In this flux model, we adjusted the albedo 
of satellite $a_{\rm mod}$ with the fixed values of the satellite temperature, $T_{\rm sat}~=~
280~{\rm K}$, and radius, $r_{\rm sat}~=~1.5~{\rm m}$, as explained in the above discussion. 

Figure 8 shows the blackbody curves with the flux of the Darksat and 
STARLINK-1113 trails in Jansky. The flux of Darksat trail is relatively bright in $g'$ 
band (April 10, 2020) or comparable in each band (May 18 and June 11, 2020). 
Meanwhile, the $g'$- and $R_{\rm c}$-band flux of STARLINK-1113 are relatively 
dimmer than that of $I_{\rm c}$ band likely due to the strong atmospheric scattering. 
It is clearly seen that the reflected radiation is dominant in the UV to optical region 
whereas the thermal radiation is dominant in the mid-infrared region 
\citep[see also][]{2020A&A...636A.121H}. By the model fitting with the blackbody radiation, 
it is confirmed that the modeled albedo of Darksat ($a_{\rm mod}=0.04$) is about a half of 
STARLINK-1113 ($a_{\rm mod}=0.075$); these values are consistent with the measured 
albedo $a_{\rm obs}$ in Table 4. Our analysis has demonstrated the effectiveness of the 
darkening treatment for Darksat. This difference of the satellite-surface albedo corresponds 
to Darksat being $\sim$ 0.75 mag darker than STARLINK-1113; this magnitude difference 
between the two satellites is consistent with the result of Tr20. Since it was difficult 
to fit the $g'$-band flux with the single-albedo blackbody model, the difference between 
$g'$-band flux and that of the other bands probably reflects the physical 
properties of the surface of these satellites.
  
\begin{figure*}
\figurenum{8}
\begin{center}
 \includegraphics[height=12cm,width=17.5cm]{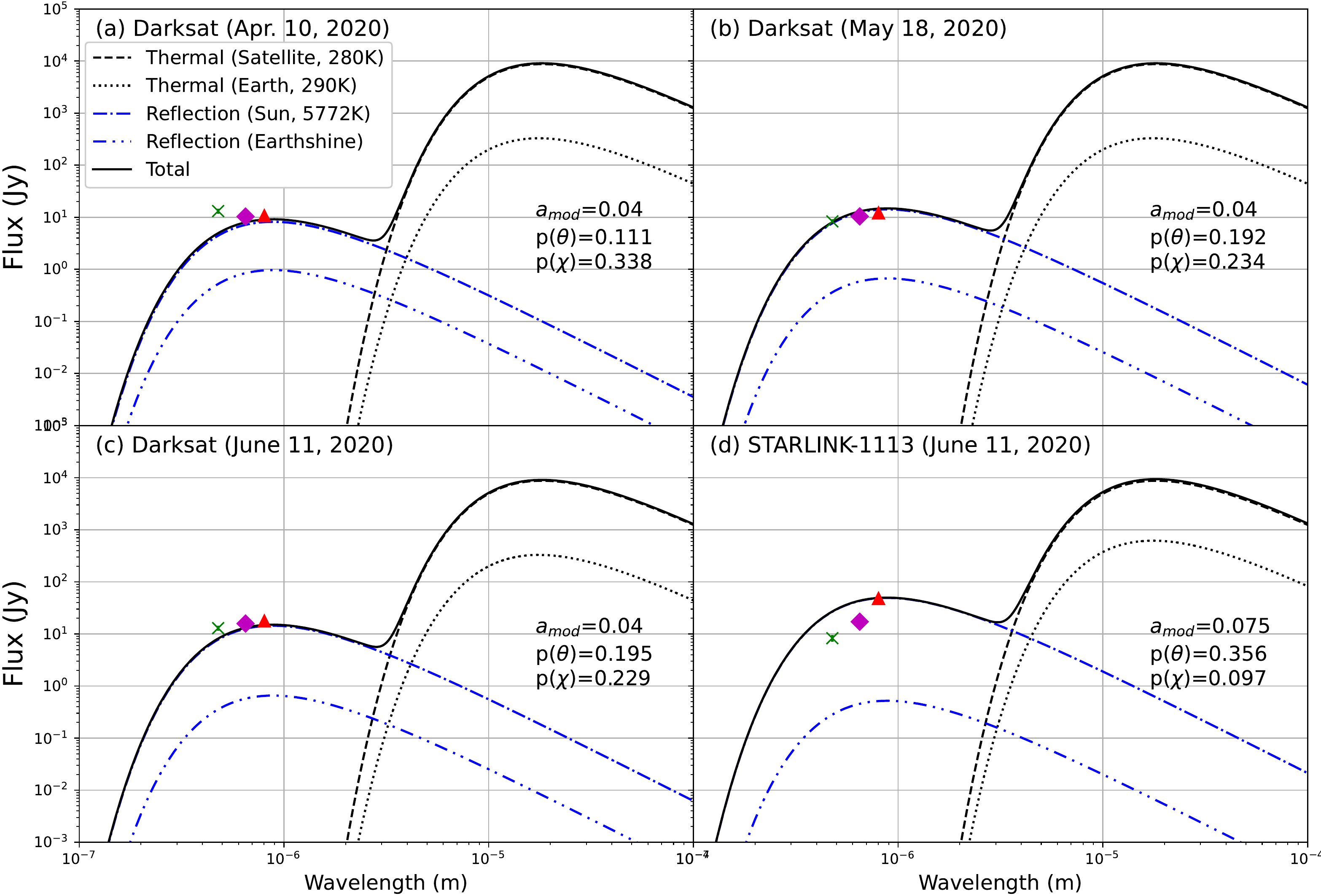} 
\caption{Modeling the $g'$- (green), $R_{\rm c}$- (magenta), and $I_{\rm c}$ (red)-band 
AB flux of Darksat (panels a, b, and c) and STARLINK-1113 (panel d) by the black-body 
radiation. In each band, the x-axis values of AB flux data points correspond to the effective 
wavelengths (see Section 2). Blue dash-dot, chain double-dashed, black dotted, dashed, 
and solid lines imply the reflection of the Sunlight, Earthshine, reflection of the Earth-thermal 
radiation, thermal radiation from these satellites, and total flux, respectively. The albedo,  
$a_{\rm mod}$, phase angle, $p(\theta)$, and $p(\chi)$ are described in each panel.  
\label{fig:f1}}
\end{center}
\end{figure*}

\section{Conclusions}
Using the 105 cm Murikabushi telescope with $\it{MITSuME}$ system, we conducted 
the simultaneous multicolor ($g'$, $R_{\rm c}$, and $I_{\rm c}$ bands) observations 
for the Starlink's LEO satellites Darksat and STARLINK-1113 on April 10, May 
18, and June 11, 2020. Our observational results are summarized as follows:   

 \begin{itemize}
      \item[(1)] The apparent $g'$-band magnitudes of Darksat are comparable to or brighter 
      than that of STARLINK-1113; 
      \item[(2)] The shorter the observed wavelength is, the fainter satellite magnitudes tend to become;  
      \item[(3)] At the vicinity of the Earth shadow, the reflected flux by STARLINK-1113 is 
      		    extremely ($>~1.0$ mag) redder than that of Darksat, excluding the difference of 
		    $g'-R_{\rm c}$; 
      \item[(4)] There is no correlation between the solar phase angle and orbital-altitude-scaled 
      		    magnitude; and 
      \item[(5)] The flux-model fitting to the satellite trails by the blackbody radiation resulted in that  
      		     the albedo of Darksat is about a half of STARLINK-1113. 
      		   
\end{itemize}

The result (1) is contrary to that of Tr20. However, our results could be explained by 
taking into account both the solar and observer phase angles and atmospheric 
extinction qualitatively. 

In June 3, 2020, Space X newly launched its eighth batch of Starlink's LEO 
satellites$\footnote{https://www.spaceflightinsider.com/missions/starlink/first-starlink-visorsat-takes-flight-aboard-spacex-falcon-9/}$. 
These satellites include STARLINK-1436 (nicknamed ``Visorsat") with a deployable Sun visor to 
prevent sunlight and reduce reflected flux. Henceforth, it is important to compare the 
astronomical impacts of the usual Starlink's LEO satellites, Darksat, and Visorsat. For precise 
verification of solar and observer phase angle dependence of satellite magnitudes, 
it is needed to carry out multicolor-multispot observation for the Starlink's LEO satellites including 
Darksat and Visorsat.  

\acknowledgments
We are grateful to the staff in IAO and Yaeyama Star Club for supporting the 
authors (TH and HH) while preparing the paper. MO is also grateful to all the staff of 
the Public Relations Center, NAOJ. We appreciate the reviewer for his/her valuable 
comments for improving this paper. The fourth U.S. Naval Observatory CCD 
Astrograph Catalog (UCAC4) is provided by Zacharias N., Finch C.T., Girard T.M., 
Henden A., Bartlet J.L., Monet D.G., Zacharias M.I. The MITSuME system was 
supported by a Grant-in-Aid for Scientific Research on Priority Areas (19047003).


\bibliography{sample63}{}
\bibliographystyle{aasjournal}

\clearpage
\appendix

\section{Approximate BRDF Model}
Contrary to the results of Tr20, the apparent and normalized magnitudes of 
STARLINK-1113 are fainter than or comparable to those of Darksat except for 
the $I_{\rm c}$-band magnitude (Figure 6). Here, we examine both the solar and observer 
phase-angle effects of Darksat and STARLINK-1113. The definition of the observer 
phase angle is the angle between the straight line from the geocenter to the satellite 
and the straight line from the observer to the satellite \citep[see the angle $c$ in Figure 1 
of][]{2020arXiv200307805M}. The previous study, Tr20, estimated a parametrised 
bidirectional reflectance distribution function (BRDF) model by \citet{1941ApJ...93..403M} 
to investigate the two phase-angle effects approximately. They evaluated the ratio, $R$, 
which is the index of solar phase attenuation between the two satellites: 
\begin{eqnarray}
R = \biggr(\frac{\cos\theta_{1113}\cos\phi_{1113}}{\cos\theta_{\rm DS}\cos\phi_{\rm DS}}\biggr)^{1-k},
\end{eqnarray}\\
where $\theta_{\rm DS}$ (or $\theta_{1113}$) and $\phi_{\rm DS}$ (or $\phi_{1113}$) 
are the solar and observer phase angle of Darksat (or STARLINK-1113), respectively. 
The Minnaert exponent $k$ ranges from 0 to 1. If the ratio $R$ is $>1$ (or $<1$), 
a phase-angle-dependent reflectance of Darksat (or STARLINK-1113) is dominant. 
Figure A shows the ratio $R(k)$ as a function of the Minnaert exponent $k$ on June 11, 2020 
with the $R(k)$ in Tr20. In this study, the phase angle component of reflected flux of Darksat is 
dominant for any $k$. Tr20 assumed a dark surface \citep[$k=0.5$;][]{1999CUP...S} 
for Darksat and STARLINK-1113. When $k=0.5$, $R(k)$ in this study will be $\sim$ 0.68 
corresponding to 0.42 mag darkening of Darksat relative to STARLINK-1113. 
Since the normalized $g'$- (or $R_{\rm c}$-) band magnitude of Darksat on June 11, 
2020 (third data points from the right in Figure 6) is already brighter than (or comparable to) 
that of STARLINK-1113, the correction by BRDF (i.e., brighten the Darksat magnitude by 0.42 
mag) leads the result that STARLINK-1113 is always darker than Darksat in $g'$ and 
$R_{\rm c}$ band. 

 \begin{figure*}
\figurenum{A}
\begin{center}
 \includegraphics[height=13cm,width=17cm]{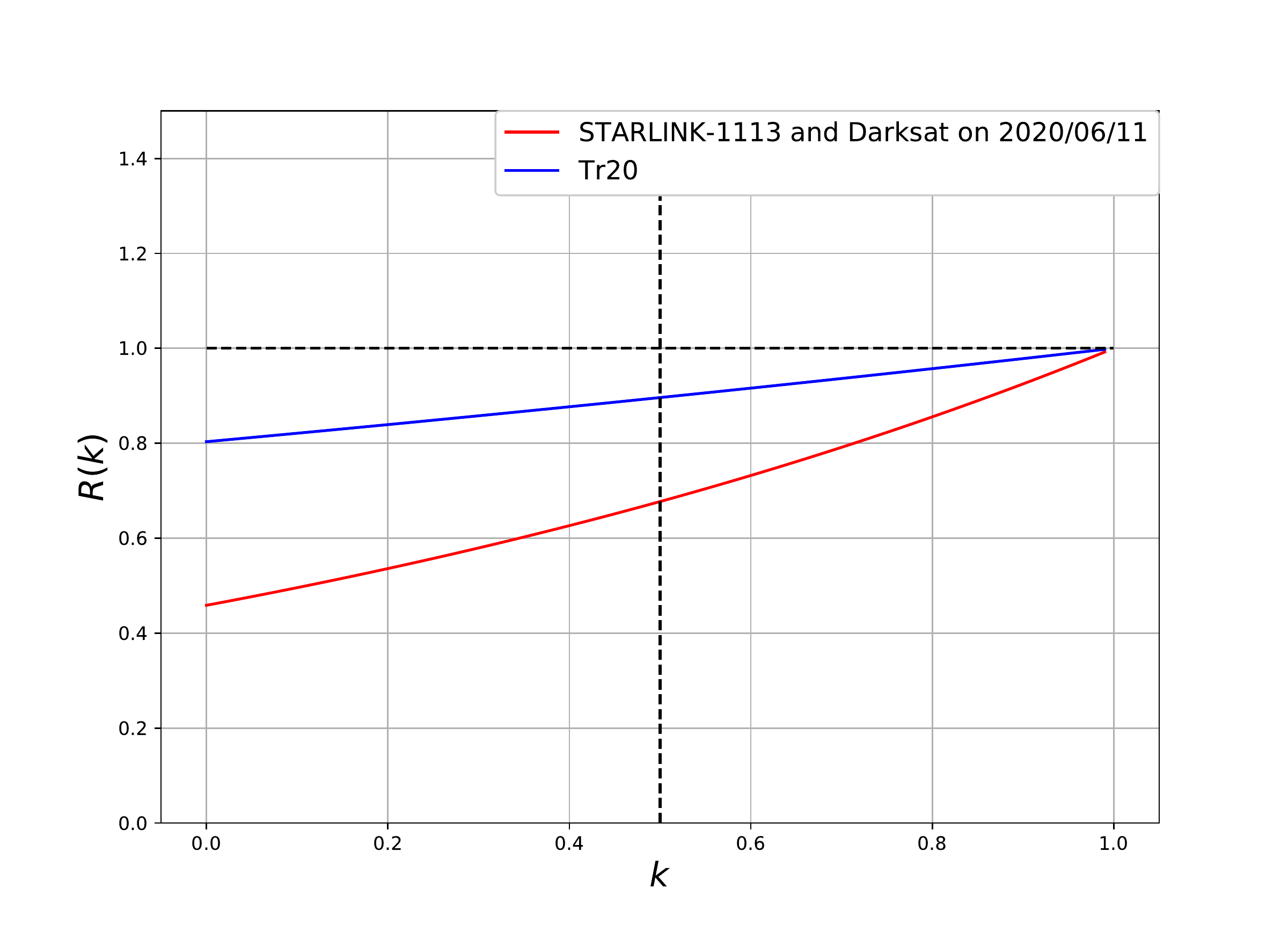} 
\caption{The ratio $R(k)$ in each epoch with the result of Tr20. 
The horizontal and vertical dashed lines imply an auxiliary line of 
$R(k)=1$ and the Minnaert exponent $k$ of 0.5 (i.e., for a dark surface), 
respectively.   
\label{fig:f1}}
\end{center}
\end{figure*}

\section{Reflection of The Earthlight and Earth Thermal Radiation}
This chapter describes the derivation for the reflection flux of the earthlight, $F_{\rm REs}$, 
and Earth thermal radiation, $F_{\rm TE}$, from the LEO satellite to an observer. The definition 
of the following parameters is the same as defined in Section 4. 

\subsection{Reflection of The Earthlight} 
The thermal radiation density of the Sun received by the Earth, $F_{\rm SE}$, 
is expressed by: 
\begin{eqnarray}
\setcounter{equation}{1}
F_{\rm SE} &=& 4\pi B(\lambda,~T_{\odot})\frac{\pi R_{\odot}^2}{4\pi(1~{\rm au})^2}~\frac{\lambda^2}{c} \nonumber \\
&=&  \pi\biggr(\frac{R_{\odot}}{1~{\rm au}}\biggr)^2B(\lambda,~T_{\odot})~\frac{\lambda^2}{c}, \\
B(\lambda,~T) &=& \frac{2hc^2}{\lambda^5}\frac{1}{\exp{(\frac{hc}{kT\lambda})}-1}.
\end{eqnarray}
As shown in Figure B, considering the distance from the Earth surface and the cross-sectional 
area of the Earth with respect to the cross-sectional area limited by the sphere with radius $l$ 
centered at the LEO satellite, the flux of the earthlight received by the LEO satellite, 
$F_{\rm ER,sat}$, is written as follows: 
\begin{eqnarray}
F_{\rm ER,sat} &=& a_{\rm E}~F_{\rm SE}~\frac{\pi (R_{\oplus}\sin{\theta_{\rm E}})^2}{\pi R_{\oplus}^2}\times
\frac{R_{\oplus}^2}{(R_{\oplus}+h_{\rm T})^2}\nonumber \\
&=& a_{\rm E}~F_{\rm SE}~\biggr(\frac{R_{\oplus}}{R_{\oplus}+h_{\rm T}}\biggr)^2\biggr\{1-\biggr(\frac{R_{\oplus}}{R_{\oplus}+h_{\rm T}}\biggr)^2\biggr\}, 
\end{eqnarray}
where $\theta_{\rm E}$ is the angle between the line from the geocenter to the 
LEO satellite and the line from the geocenter to the interacting point on the ground 
surface with the tangent line between the LEO satellite and ground surface. 
As a result, the earthlight flux from the LEO satellite received by an observer, 
$F_{\rm REs}$, is described by: 

\begin{eqnarray}
F_{\rm REs} &=& F_{\rm ER,sat}~a_{\rm mod}\biggr(\frac{r_{\rm sat}}{r}\biggr)^2~p(\chi)~
\biggr(\frac{r}{h_{\rm T}}\biggr)^2 \nonumber \\
&=& \pi\biggr(\frac{R_{\odot}}{1~{\rm au}}\biggr)^2B(\lambda,~T_{\odot})a_{\rm E}\biggr(\frac{R_{\oplus}}{R_{\oplus}+h_{\rm T}}\biggr)^2\biggr\{1-\biggr(\frac{R_{\oplus}}{R_{\oplus}+h_{\rm T}}\biggr)^2\biggr\}a_{\rm mod}\biggr(\frac{r_{\rm sat}}{h_{\rm T}}\biggr)^2~p(\chi)~\frac{\lambda^2}{c}\nonumber \\
&=& a_{\rm E}\biggr(\frac{R_{\oplus}}{R_{\oplus}+h_{\rm T}}\biggr)^2\biggr\{1-\biggr(\frac{R_{\oplus}}{R_{\oplus}+h_{\rm T}}\biggr)^2\biggr\}~
\frac{p(\chi)}{p(\theta)}~F_{\rm RS}, 
\end{eqnarray}
where this flux is normalized with the orbital height, $h_{\rm T}$, by the factor, 
$(r/h_{\rm T})^2$. 

\subsection{Reflection of The Earth thermal radiation}
The thermal radiation density of the Earth received by the LEO satellite, $F_{\rm ET,sat}$, 
is evaluated as follows: 

\begin{eqnarray}
F_{\rm ET,sat} &=& 4\pi\epsilon~B(\lambda, T_{\rm E})~
\frac{\pi (R_{\oplus}\sin{\theta_{\rm E}})^2}{4\pi l^2}~\frac{\lambda^2}{c}\nonumber \\
&=& \pi \epsilon \biggr(\frac{R_{\oplus}}{R_{\oplus}+h_{\rm T}}\biggr)^2~B(\lambda, T_{\rm E})~\frac{\lambda^2}{c}. 
\end{eqnarray}
Namely, the Earth thermal radiation from the LEO satellite received by an observer, 
$F_{\rm TE}$, normalized with the orbital height is expressed by: 

\begin{eqnarray}
F_{\rm TE} &=& F_{\rm ET,sat}~a_{\rm mod}\biggr(\frac{r}{h_{\rm T}}\biggr)^2 \nonumber \\
&=& \pi\epsilon\biggr(\frac{R_{\oplus}}{R_{\oplus}+h_{\rm T}}\biggr)^2B(\lambda,~T_{\rm E})
~a_{\rm mod}\biggr(\frac{r_{\rm sat}}{h_{\rm T}}\biggr)^2~\frac{\lambda^2}{c}. 
\end{eqnarray}

\begin{figure*}
\figurenum{B}
\begin{center}
 \includegraphics[height=8.5cm,width=12cm]{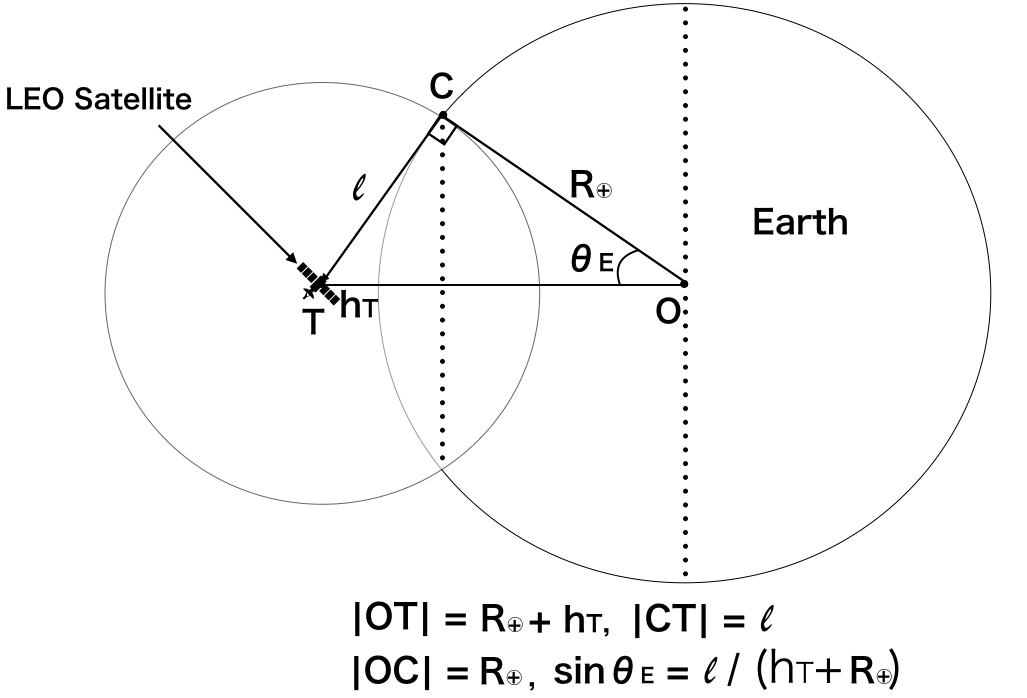} 
\caption{The cross-sectional area of the Earth with respect to the cross-sectional 
area limited by the sphere centered at the LEO satellite. Where $\theta_{\rm E}$ is 
the angle between the line from the geocenter (Point O) to the the LEO satellite (Point T) 
and the line from the geocenter to the interacting point on the ground surface
with the tangent line (length $l$) between the LEO satellite and ground 
surface (Point C).}
\end{center}
\end{figure*} 

\section{Thermal- and Reflected-Satellite Flux by The Moon}
In Section 4, we discussed the normalized thermal and reflected flux of Darksat 
and STARLINK-1113 without the effect by the Moon. The normalized thermal, 
$F_{\rm TM}$, and reflected flux, $F_{\rm RM}$, of the satellites by the Moon 
are described with: 
\begin{eqnarray}
\setcounter{equation}{1}
F_{\rm TM} &=& \pi\epsilon\biggr(\frac{R_{\rm Moon}}{D_{\rm Moon}}\biggr)^2B(\lambda,~T_{\rm Moon})
~a~p(\theta)~\biggr(\frac{r_{\rm sat}}{h_{\rm T}}\biggr)^2~\frac{\lambda^2}{c}, \\
F_{\rm RM} &=& \pi\biggr(\frac{R_{\odot}}{1{\rm au}}\biggr)^2B(\lambda,~T_{\odot})~a_{\rm Moon}~
p(\theta_{\rm Moon})\biggr(\frac{R_{\rm Moon}}{D_{\rm  Moon}}\biggr)^2~a~p(\theta)~
\biggr(\frac{r_{\rm sat}}{h_{\rm T}}\biggr)^2~\frac{\lambda^2}{c},
\end{eqnarray}
where $R_{\rm Moon}$ (= $1.73~\times~10^3$ km), $T_{\rm Moon}$, $a_{\rm Moon}$ (= 0.12), 
$\theta_{\rm Moon}$ (or $\theta$), $p(\theta_{\rm Moon})$ (or $p(\theta)$), and 
$D_{\rm Moon}$ (= $3.83~\times~10^5$ km) are the Moon radius, irradiated-Moon-surface 
temperature, albedo of the Moon, Sun-Moon-satellite (or Moon-satellite-observer) phase angle, 
phase integral, and the distance between an observer and the Moon, respectively. The definition 
of the other parameters in Equations (C1) and (C2) is the same as defined in Section 4. 



\end{document}